\documentclass[aoas]{imsart}

\RequirePackage{amsthm,amsmath,amsfonts,amssymb}
\RequirePackage[authoryear]{natbib}

\usepackage{url}
\usepackage{graphicx}
\startlocaldefs

\newcommand\bzero{\mbox{\boldmath${0}$}}
\newcommand\bbe{\mbox{\boldmath${ \beta}$}}

\newcommand\bfeta{\mbox{\boldmath${\eta}$}}
\newcommand\bep{\mbox{\boldmath${\epsilon}$}}
\newcommand\bga{\mbox{\boldmath${\gamma}$}}

\newcommand\bdel{\mbox{\boldmath${\delta}$}}

\newcommand\bxi{\mbox{\boldmath${\xi}$}}

\newcommand\bSig{\mbox{\boldmath${\Sigma}$}}

\newcommand\btheta{\mbox{\boldmath${\theta}$}}

\newcommand\bI{{\bf I}}

\newcommand\mcN{{\mathcal N}}

\newcommand\mcS{{\mathcal S}}
\newcommand\bs{{\bf s}}

\newcommand\bX{{\bf X}}

\newcommand\bY{{\bf Y}}

\newcommand\bZ{{\bf Z}}

\newcommand\mbR{{\mathbb R}}

\endlocaldefs

\begin{document}

\begin{frontmatter}
\title{Probabilistic Downscaling for Flood Hazard Models}
%\title{A sample article title with some additional note\thanksref{T1}}
\runtitle{Probabilistic Downscaling for Flood Hazard Models}
%\thankstext{T1}{A sample of additional note to the title.}

\begin{aug}
%%%%%%%%%%%%%%%%%%%%%%%%%%%%%%%%%%%%%%%%%%%%%%%
%% Only one address is permitted per author. %%
%% Only division, organization and e-mail is %%
%% included in the address.                  %%
%% Additional information can be included in %%
\author[A]{\fnms{Samantha}~\snm{Roth}},
\author[B]{\fnms{Sanjib}~\snm{Sharma}},
\author[C]{\fnms{Atieh}~\snm{Alipour}},
\author[A]{\fnms{Klaus}~\snm{Keller}},
\and
\author[D]{\fnms{Murali}~\snm{Haran}}
%%%%%%%%%%%%%%%%%%%%%%%%%%%%%%%%%%%%%%%%%%%%%%
%% Addresses                                %%
%%%%%%%%%%%%%%%%%%%%%%%%%%%%%%%%%%%%%%%%%%%%%%
\address[A]{Thayer School of Engineering at Dartmouth College}

\address[B]{Department of Civil and Environmental Engineering, Howard University}

\address[C]{National Oceanic and Atmospheric Administration}

\address[D]{Department of Statistics, The Pennsylvania State University}
\end{aug}

\begin{abstract}
%Riverine flooding drives sizable risks. Designing strategies to manage flood risks hinges on flood projections with decision-relevant scales and uncertainties. Many decisions require high spatial resolutions. Calibrating high resolution flood models can impose prohibitive computational demands. One approach to tackle this challenge is to downscale a low resolution model projection by mapping it onto a higher resolution grid. Here we develop and test a probabilistic downscaling approach for flood hazard models.  

Riverine flooding poses significant risks. Developing strategies to manage flood risks requires flood projections with decision-relevant scales and well-characterized uncertainties, often at high spatial resolutions. However, calibrating high-resolution flood models can be computationally prohibitive. To address this challenge, we propose a probabilistic downscaling approach that maps low-resolution model projections onto higher-resolution grids.
%In the existing literature there are (1) probabilistic downscaling approaches that can be applied to many types of physics-based models and (2) deterministic downscaling approaches that are developed specifically for flood hazard models. 
The existing literature presents two distinct types of downscaling approaches: (1) probabilistic methods, which are versatile and applicable across various physics-based models, and (2) deterministic downscaling methods, specifically tailored for flood hazard models. Both types of downscaling approaches come with their own set of mutually exclusive advantages. 
Here we introduce a new approach, PDFlood, that combines the advantages of existing probabilistic and flood model-specific downscaling approaches, mainly (1) spatial flooding probabilities and (2) improved accuracy from approximating physical processes. Compared to the state of the art deterministic downscaling approach for flood hazard models, PDFlood allows users to consider previously neglected uncertainties while providing comparable accuracy, thereby better informing the design of risk management strategies. While we develop PDFlood for flood models, the general concepts translate to other applications such as wildfire models. 
\end{abstract}


\end{frontmatter}

%%%%%%%%%%%%%%%%%%%%%%%%%%%%%%%%%%%%%%%%%%%%%%
%% Please use \tableofcontents for articles %%
%% with 50 pages and more                   %%
%%%%%%%%%%%%%%%%%%%%%%%%%%%%%%%%%%%%%%%%%%%%%%
%\tableofcontents

%%%%%%%%%%%%%%%%%%%%%%%%%%%%%%%%%%%%%%%%%%%%%%
%%%% Main text entry area:

%%%%%%%%%%%%%%%%%%%%%%%%%%%%%%%%%%%%%%%%%%%%%%
%% Appendix---Please move all appendices to %%
%% a Supplementary file.                    %%
%%%%%%%%%%%%%%%%%%%%%%%%%%%%%%%%%%%%%%%%%%%%%%
%% Support information, if any,             %%
%% should be provided in the                %%
%% Acknowledgements section.                %%
%%%%%%%%%%%%%%%%%%%%%%%%%%%%%%%%%%%%%%%%%%%%%%

\section{Introduction}\label{Sec: Intro}

Riverine flooding drives sizable risks for human life and property. Compared to other natural disasters, flooding impacts the greatest number of people \citep{CRED2015}. These risks are expected to increase with urbanization and climate change \citep{easterling_2017,Dottori_etal_2018}.
%Modeled results indicate the economic impacts of flooding can be widespread https://www.tandfonline.com/doi/full/10.1080/09535314.2016.1232701#d1e2466

Decision-makers seek flood hazard projections to inform efforts to reduce flood risk \citep{HomeownersGuideToRetrofitting}. Flood risk is the product of hazard, exposure, and vulnerability \citep{managingRisks}. One common definition for flood hazard in a given location is the maximum depth of flood water during a flood event \citep{merz_etal_2010}. The people and property that may be impacted by hazards comprise exposure \citep{LivingWithRisk}. Vulnerability describes the propensity of exposed elements to be damaged by hazards \citep{LivingWithRisk}. Increasing hazards can motivate actions to reduce exposure and vulnerability \citep{Simpson_etal_2021,Siders_2019}. For example, to reduce exposure, the government may buy back properties that flood too frequently. To reduce vulnerability, a homeowner may elevate, wet-proof, or dry-proof their home \citep{HomeownersGuideToRetrofitting}. Stakeholders may consider the property-level projected flood hazards to guide these choices.

Flood hazard projections frequently lack the appropriate spatial detail to inform property-level decisions. Exposure can vary on scales as small as 5 meters (m) \citep{PhillyZoningCode}. Private firms often provide national and global-scale flood hazard maps at 30 m resolution \citep[e.g.][]{FSF,Fathom}. One grid cell of these flood projections could be as wide as several properties \citep{PhillyZoningCode}. While these flood projections have higher spatial resolution than previous national and global-scale flood hazard products, they are still very coarse for urban areas \citep{fewtrell_etal_2008}. 

Downscaling is a method for mapping outputs of physics-based models to the appropriate scale. For many physics-based models, including riverine flood hazard models, producing projections at the desired high resolution directly is very challenging given the current computational resources. By downscaling low resolution projections, researchers can obtain an approximation of these high resolution projections. Downscaling approaches applied to riverine flood hazard models can be broadly categorized as deterministic or statistical.

%statistical downscaling approaches have feature 

Statistical downscaling approaches can provide uncertainty quantification and use information from observational data but typically neglect specific mechanisms for downscaling flood hazards \citep[e.g.][]{fuentesAndRaftery2005,wikleAndBerliner2005,berrocal_etal_2010}. These approaches can fit a statistical relationship between low resolution model projections and observations \citep[e.g.][]{berrocal_etal_2010} to naturally provide uncertainty quantification. Observational data can provide valuable insights to the downscaling process. However, some statistical downscaling approaches assume spatially correlated model parameters \citep[e.g.][]{berrocal_etal_2010,wikleAndBerliner2005}, requiring more observations than are typically available in flood hazard modeling \citep{Parkes_etal_2013}. 

Many recently developed statistical downscaling approaches employ deep learning methods \citep[e.g.][]{Carreau_Guinot_2021,wu_etal_2021,francois_etal_2021}. However, these approaches require training using a representative set of high resolution model runs, rendering them inappropriate for situations in which the model cannot be run at the desired high resolution. We consider this data-poor situation which is typical of riverine flood hazard modeling. To our knowledge, existing statistical downscaling approaches have not addressed this challenging situation. Finally, because many statistical downscaling approaches are developed to apply to many types of physics-based models, they also neglect useful procedures specific to downscaling flood hazards.
%Examples of more recent statistical downscaling approaches applied to flood hazards neglect uncertainty quantification and rely on access to a representative sample of high resolution model runs, which is not always available \citep[e.g.][]{Carreau_Guinot_2021,Carreau_Naveau_2023}.
%berrocal et al: regression parameters modeled as having spatially varying part and non-spatially varying part, where each part in turn also follows a distribution
%wikle et al: model a spatially correlated “true” unobserved process that is related to the observations via measurement error models

%do the possibly flood model specific approaches I found provide any UQ?

Hydrologists have developed downscaling approaches specifically for riverine flood projections, but these approaches are typically silent on questions surrounding uncertainty quantification \citep[e.g.][]{Schumann_2014,bryant_etal_2024}. These approaches leverage high resolution elevation data and have separate schemes for flooded and non-flooded low resolution cells \citep{Schumann_2014,bryant_etal_2024}. One prominent example, CostGrow, employs cost distance analysis to approximate the spread of flood water outside the low resolution flooded area \citep{bryant_etal_2024}. This approach achieves a higher degree of accuracy compared to simpler approaches without flood model-specific schemes \citep{bryant_etal_2024}. However, this flood model-specific approach is deterministic and neglects uncertainty in its approximation of high resolution flood hazard projections.

Here we propose a probabilistic downscaling approach developed specifically for flood hazard models that merges the advantages of both types of downscaling approaches, which we call ``PDFlood.'' Like some statistical downscaling approaches \citep[e.g.][]{wikleAndBerliner2005,berrocal_etal_2010}, PDFlood provides model-based uncertainty quantification and uses observational data. PDFlood's simpler model structure compared to other statistical downscaling approaches allows for easier model-fitting using scarce observed flood heights and no high resolution flood projections. Like other flood model-specific methods, PDFlood uses high resolution elevation data, employs different approaches for flooded and non-flooded low resolution cells, and makes use of cost distance analysis. By combining statistical downscaling with flood model-specific downscaling, PDFlood achieves comparable accuracy to CostGrow while quantifying neglected uncertainties. PDFlood allows decision-makers who may wish to reduce exposure or vulnerability to consider a prediction interval of flood heights in addition to the point estimates provided by a traditional downscaling approach.

%MH: ADD 1-2 SENTENCES ON HOW PDFLOOD IS BETTER FOR DECISION-MAKING? 

\section{Methods}\label{Sec: Probabilistic Downscaling Method}

\subsection{Case study details}\label{SS: Case Study Details}

We demonstrate our approach for a case study in Norristown, a suburb of Philadelphia (PA). Norristown is a municipality along the Schuylkill River that has experienced repeated riverine flooding \citep{NorristownFloodHistory}. In urbanized areas like Norristown both buildings and the distances between them are sometimes only a few meters wide. Spatially coarse hazard characterizations that miss these details can miss important risks and dynamics. 

We use LISFLOOD-FP, a two-dimensional hydraulic model, to produce flood hazard projections. Specifically, we use the subgrid formulation to simulate maximum flood depths \citep{lisflood}. LISFLOOD-FP can project flooding at a wide range of spatial and temporal scales \citep[e.g.][]{regionalFloodRisk,Rajib_etal_2020,LisfloodCongo}. LISFLOOD-FP uses a finite difference scheme on a staggered grid to solve the local inertial form of the shallow water equations. The model uses inputs related to ground elevation data describing the floodplain topography, channel bathymetry information (river width, depth, and shape), and inflow to the modeling domain as the boundary condition information. To apply LISFLOOD-FP, we use the subgrid-scale hydrodynamic scheme of \citet{nealetal_2012} to solve the momentum and continuity equations for both channel and floodplain flow. The scheme operates on a rectangular grid mesh of the same resolution as the input elevation map, i.e. digital elevation model (DEM). The cells’ water depths are updated using mass fluxes between cells, ensuring mass conservation. 

Flood hazard projections depend on often uncertain inputs. We refer to unknown inputs as parameters. We calibrate an unknown parameter to produce flood hazard projections informed by observations \citep{kennedy2001bayesian}. We use five observations of maximum flood depth, called high water marks, from Hurricane Ida (2021) for calibration \citep{USGS_HWMs}. We consider Hurricane Ida due to the availability of observations and the extensive damages \citep{USGS_Ida}. We focus on a single parameter, Manning's roughness coefficient for the channel ($n_{ch}$). This parameter approximates the resistance to flow in the channel \citep{arcement1989}. Flood models are generally very sensitive to channel roughness \citep{pappenberger_2008,savage_etal2016,alipour_etal2022}. We set the range of plausible values for channel roughness as $n_{ch} \in (0.01,0.1)$ based on prior work 
\citep{alipour_etal2022, pappenberger_2008}. We treat the DEM and other floodplain topography data as known. We obtain the floodplain topography data from the Pennsylvania Spatial Data Access (PASDA) archive \citep{PASDA}. To set the resolution of the flood hazard projections, we use 5 m and 10 m DEMs. We aggregate these digital elevation models from a 1 m resolution DEM constructed from PAMAP LiDAR (Light Detection and Ranging) elevation points \citep{PAMAP_LIDAR_1mDEM}. For a 5 m DEM, the size of a grid cell is $5 \times 5$ $\textnormal{m}^2$. 

Higher spatial resolution flood hazard projections tend to be more computationally expensive (Table \ref{Table:TimeToGetRuns}). At the lower spatial resolution considered, 10 m, a single model run has a mean wall time of 5.9 minutes on a single core of the Pennsylvania State University's ICDS Roar supercomputer. At the 10 m spatial resolution a single model run can take as few as 2 minutes or as many as 8.9 minutes. At the higher spatial resolution, 5 m, a single model run  has an average wall time of 2 hours. However, a 5 m resolution model run can take as short as 24 minutes or as long as 7 hours. The time to obtain a single model run at a fixed spatial resolution varies depending on the channel roughness value. We use PDFlood to downscale a 10 m resolution projection to the 5 m resolution. We evaluate PDFlood's ability to approximate high resolution flood projections for three recent flood events and one hypothetical flood event (Table \ref{Table:Floods}). To obtain a projection for each flood event we use the corresponding river discharge corresponding as the inflow boundary condition.

\begin{table}[t]
\caption{Wall time to produce projections at each considered resolution}
\centering
\begin{tabular}{rllll}
  \hline
  & 10 m & 5 m \\ 
  \hline
  Min & 2 minutes& 24 minutes \\
  Mean & 5.9 minutes & 2 hours\\
  Max & 8.9 minutes & 7 hours\\
  \hline
\end{tabular}
\label{Table:TimeToGetRuns}
\end{table}

\begin{table}[t]
\caption{Flood events used to test PDFlood \citep{USGS_peakflow}.}
\centering
\begin{tabular}{rllll}
  \hline
  Flood event & River discharge ($\frac{m^3}{s}$)\\ 
  \hline
  Record-setting rainfall (2014) & 2560 \\
  Tropical Storm Isaias (2020) & 2503 \\
  Hurricane Ida (2021) & 3092 \\
  Hypothetical & 3681 \\
  \hline
\end{tabular}
\label{Table:Floods}
\end{table}

\subsection{Bayesian emulation-calibration}\label{SS: Emulation Calibration}

First we use calibration to select the most realistic flood hazard projection. 
We produce this projection using the most realistic combination of parameter values. Here we infer the most realistic value of $n_{ch}$. %MH: I THINK YOU CAN DELETE THE NEXT TWO SENTENCES AND BEGIN WITH THE SENTENCE ABOUT THE BAYESIAN APPROACH; NO NEED TO START THAT ONE ON A NEW PARAGRAPH. Calibrating the parameter requires choosing a measure of distance such as mean-squared error between the observations and the projections to evaluate how realistic different parameter values are. We also must decide whether to treat the parameters as fixed or random. 
We adopt a Bayesian approach which treats the model parameters as random. The Bayesian calibration framework accounts for (1) expert judgments about parameters prior to seeing the observations, (2) observational errors, and (3) systematic discrepancies between the model projections and observations \citep[e.g.][]{kennedy2001bayesian}. %Deterministic calibration treats model parameters as fixed and neglects observational errors and systematic model-observation discrepancies. Using a deterministic model calibration approach leads to an increased chance of overfitting your computer model to the observational data \citep{VivekAndKlausABM}. For these reasons
We use Bayesian calibration to obtain a calibrated projection at each of the 10 m and 5 m resolutions. The calibrated 5 m resolution projection mimics the observational data relatively well considering the scale at which observations are gathered. Please see the supplement for details about Bayesian calibration and the calibrated 5 m projection performance.

%In Bayesian calibration, we assign probability distributions to the model parameters to express our beliefs about their values before seeing the observations, $\pi(\theta)$. We also assign a probability distribution $\pi(z|y)$ to express the relationship between the model output $y$ and the observational data $z$. Here, the relationship between the model parameters and output is deterministic, so we can replace $\pi(z|y)$ with  $\pi(z|\theta)$. We then use these relationships to infer the probability distribution $\pi(\theta|z)$ of $\theta$ after observing $z$:
%$$\pi(\theta|z)\propto \pi(\theta) \pi(z|\theta).$$

%I use the posterior median of $n_{ch}$ to produce the calibrated projection in this case. So then that would be the flood projection that minimizes mean absolute error, accounting for the model-observation discrepancy. That feels like a mouthful that might not be appropriate for a signal sentence so I'm a bit stuck.

%\subsection{Overview of probabilistic downscaling for flood models}

%We specify different probability distributions for high resolution flood heights in different locations. 

%The probability distribution of high resolution flood heights also depends on the low resolution flood heights, high and low resolution elevations, and observed flooding.

\subsection{Cost distance analysis}\label{SS: Cost Distance Analysis}
Cost distance analysis has been used previously in flood model-specific downscaling \citep{bryant_etal_2024} and in fast approximations to flood hazard models \citep[e.g.][]{williams_and_luckvogel_2020,Debbarma_etal_2024}. Cost distance analysis determines the least costly route for travel between two points in two-dimensional space. The definition of cost is problem-dependent and can quantify limiting factors such as physical effort or money \citep[e.g.][]{Foltete_etal_2008,collischonn&Pilar2000}. The least-cost path algorithm developed by \citet{Dijkstra1959} set the foundation for cost distance analysis. Cost distance analysis takes as inputs (1) a grid with cells labeled either as starting points (sources) or end points (destinations) and (2) a cost-of-passage grid. A cost-of-passage grid assigns to each cell a cost of passing through \citep{collischonn&Pilar2000}. For each destination cell, cost distance analysis computes (1) the least-cost path starting from any source cell, (2) the source cell whose path costs the least, and (3) the accumulated cost of traveling from the least-cost source cell to the destination cell \citep{collischonn&Pilar2000}. %The accumulated cost grid gives this accumulated cost at each destination cell . 

%maybe I can get rid of the middle two paragraphs?

The cost-of-passage for each cell characterizes the difficulty of movement through the cell. Factors influencing cost-of-passage may include surface type and what is passing through the cell. In our case flood water is passing through. Cost-of-passage can depend on the direction of travel through the cell. If this is the case, then a cost-of-passage grid for each direction of travel may be appropriate. Direction of travel may influence cost-of-passage due to factors like slope and wind direction.

We can use the grid of cost-of-passage cells to calculate the total cost of travel starting from any cell on the grid and finishing at any cell on the grid. For example, assume there are N cells to travel through to get to a given destination cell with cell 1 being the source cell and cell N being the destination cell. Let $c_{i,j}$ be the cost of moving from cell $i$ to cell $j$ given that no other cells must be crossed to get from cell $i$ to $j$. Then the accumulated cost $C_{1,N}$ at the destination cell is:
$$C_{1,N}=\sum_{n=1}^{N-1} c_{n,n+1}.$$ 
We assume that for two adjacent cells, $c_{n,n+1}= \frac{c_n + c_{n+1}}{2}$ and for two cells diagonal from each other $c_{n,n+1}= \frac{\sqrt{2}(c_n + c_{n+1})}{2}$. Here we assume direction of travel does not matter and $c_i$ is the cost-of-passage through cell $i$. We perform cost distance analysis using the WhiteboxTools plugin \citep{WhiteboxTools} in QGIS \citep{QGIS}.

Using cost distance analysis we approximate how water flows between grid cells. We do this to learn (1) which high resolution cells outside the low resolution flooded area may be flooded and (2) how much they may be flooded. We let $c_i$ equal the elevation of cell $i$ to account for the difficulty of flood water reaching higher elevations. We define the source cells as high resolution grid cells within the low resolution flooded area and the destination cells as the high resolution cells outside the low resolution flooded area. We then identify the source cell with the least-cost path for each destination cell. We use the source cell with the least-cost path to inform the probability distribution of flood heights for each destination cell. We provide more information on this procedure in Subsection 2.5.

\subsection{Model for flooded low resolution cells}\label{SS: Model for wet low resolution cells}

Adapting the different treatments of flooded and non-flooded low resolution cells in the deterministic flood model-specific downscaling literature \citep{bryant_etal_2024, Schumann_2014}, we specify different probability models to downscale flooded and non-flooded low resolution cells. If the high resolution grid cell is within a flooded low resolution cell, we model its flood height with a t distribution. Normal distributions are commonly used to model flood heights \citep[e.g.][]{hall_etal_2011,me2023,sanjib_etal_2023}. Instead, because we use a small sample of observational data to estimate the variance, we use a t distribution to model unknown high resolution flood heights. By scaling and shifting the t distribution, we can easily define the expected flood height and quantify the surrounding uncertainty using the mean, variance, and degrees of freedom parameters. Because the t distribution has support everywhere and flood heights cannot be negative, we treat any negative predicted values as zero. We find that this approach works well in practice.

First we estimate the mean using the low resolution flood height and the low and high resolution elevations at the current high resolution grid cell. Let $v$ be the current high resolution grid cell. Let $Y_H(v)$ and $E_H(v)$ be the high resolution flood height and elevation at $v$. Let $(w_1,w_2,w_3,w_4)$ be the four closest low resolution grid cells. Let $(Y_L(w_1),Y_L(w_2),Y_L(w_3),Y_L(w_4))$ and $(E_L(w_1),E_L(w_2),E_L(w_3),E_L(w_4))$ be the low resolution flood heights and elevations at these locations. Following \citet{bryant_etal_2024}, we obtain a downscaled flood height $Y_D(v)$ in two steps: 
\begin{enumerate}
    \item Bilinearly interpolate $$(Y_L(w_1) + E_L(w_1),Y_L(w_2) + E_L(w_2),Y_L(w_3) + E_L(w_3),Y_L(w_4) + E_L(w_4))$$ onto $v$ to obtain $Y_B(v)$.
    \item Let $Y_D(v) = [Y_B(v) - E_H(v)]_+$.
    %\item The result is the downscaled flood height $Y_D(v) = $. If negative, set $Y_D(v)$ to zero. 
\end{enumerate}
Here $[\cdot]_+=\textnormal{max}(\cdot,0)$, i.e. $[\cdot]_+$ sets $\cdot$ equal to zero if negative. We set the mean of our predictive distribution for $Y_H(v)$ equal to $Y_D(v)$. 

We estimate the variance using the observed flood heights and the distribution mean. We start by assuming $Y_H(v) \sim \mcN(Y_D(v),\sigma^2)$. %$$\frac{Z(u_i)-Y_D(u_i)}{\sigma} \sim \mcN(0,1).$$
Since we do not know $\sigma^2$, the predictive distribution for $Y_H(v_0)$ at a new location $v_0$ would be $$Y_H(v_0) \sim Y_D(v_0) + \hat{\sigma} \mathcal{T}(n-1).$$ Here $\mathcal{T}(n-1)$ is the t distribution with $n-1$ degrees of freedom, and $n$ is the number of high resolution model runs used to compute $\hat{\sigma}$, the unbiased estimator for $\sigma$. We shift and scale the t distribution by $Y_D(v_0)$ and $\hat{\sigma}$, respectively. However, we do not have access to any high resolution model runs to compute $\hat{\sigma}$, so all $v$ are new locations. To work around this challenge, we replace the high resolution model runs with observations $\bZ = (Z(u_1),...,Z(u_5))$ in our computation of $\hat{\sigma}^2$, where $((u_1),...,(u_5))$ are the observation locations. %as follows: $$\hat{\sigma}^2 = \frac{\Sigma_{i=1}^{5}(Z(u_i) - Y_D(u_i))}{4}.$$ Here $Z(u_i)$ is the observed flood height at location $u_i$.
Since the number of observations used to estimate $\hat{\sigma}^2$ is small, we do not assume that the t distribution can be reasonably approximated by a normal distribution.

%Because we estimate $\sigma$ from the observations, the predictive distribution for $Z$ at a new location $u_0$ becomes $\frac{Z(u_0)-Y_D(u_0)}{\hat{\sigma}} \sim \mathcal{T}(n-1)$, where $n=5$ is the number of observations. 
%First we obtain the downscaled flood heights $\bY_D(\bu)= (Y_D(u_1),...,Y_D(u_{5}))$ at the five observation locations $\bu=(u_1,...,u_{5})$. We then approximate the model variance as  $$\hat{\sigma}^2 = \frac{\Sigma_{i=1}^{5}(Z(u_i) - Y_D(u_i))}{4},$$
%where $Z(u_i)$ is the observed flood height at location $u_i$. $\hat{\sigma}^2$ is an unbiased estimator of $\sigma^2$ for the model $\bZ \sim \bY_D(\bu) + \hat{\sigma}^2 \mathcal{T}(\nu)$. Our model is $$\bY_H(\bv) \sim \bY_D(\bv) + \sigma^2 \mathcal{T}(\nu),$$ where $\bv$ is the collection of all high resolution locations. $\mathcal{T}(\nu)$ is the t distribution with degrees of freedom $\nu$, where here $\nu=4$. However, we do not have access to entries of $\bY_H(\bv)$ to use to estimate $\sigma^2$. If we did, our probabilistic downscaling approach would serve no purpose. 

This approximation is motivated by our belief that differences between observational data and downscaled flood heights should be more variable than the differences between flood heights from a high resolution version of the same physics-based model and our downscaled flood heights. Some discrepancies between physics-based models and reality do not resolve by increasing the spatial resolution \citep[see for example the discussion in][]{ReifiedBayesianModelling}. However, if the spatial resolution we aim to approximate is high and our physics-based model is a reasonable representation of the physical system it approximates, then our approximation $\hat{\sigma}^2$ should provide value in estimating $\sigma^2$.

\subsection{Model for dry low resolution cells}\label{SS: Model for dry low resolution cells}

If the high resolution grid cell $v$ is within a non-flooded low resolution cell, we model its flood height with a mixture of a shifted and scaled t distribution $Y_A(v) + \hat{\sigma}\mathcal{T}(n-1)$ and a point mass at zero $I(0)$. Our model is:
$$ Y_H(v) | c \sim (1-\pi)I(0) + \pi (Y_A(v) + \hat{\sigma}\mathcal{T}(n-1)).$$
The weights for the mixture, $\pi$ and $1-\pi$, are the probabilities that $Y_H(v)$ comes from $Y_A(v) + \hat{\sigma}\mathcal{T}(n-1)$ versus $I(0)$, respectively. $Y_A(v)$ is an adjusted downscaled flood height at location $v$.

The weights in the mixture distribution exploit the relationship between elevation and flood presence. See Figure \ref{Fig: elev vs flood prob} in the supplement for illustration of the inverse relationship between elevation and flooding probability. We assume that the probability $\pi$ of a cell $v$ being flooded given elevation $E(v)$ is the same no matter the resolution of the cell. Figure \ref{Fig: elev vs flood prob} also shows that this assumption is reasonable. Because we do not have access to the high resolution flood projection, we estimate the probability of flooding at a given elevation using the low resolution flood projection. In our estimation process, we first sort the low resolution cells into groups with similar elevations. We then use a Gaussian process to interpolate the proportion of low resolution cells that are flooded between the midpoints of the elevation groups. We define $\pi$ to be the conditional mean function of this Gaussian process given the selected low resolution projection and elevations. See the supplement for additional details surrounding this process.

%Use cost distance analysis to identify the nearest high resolution cell in a flooded low resolution cell.
We estimate the remaining parameters of the mixture distribution using information extracted using cost distance analysis. First we use cost distance analysis to identify where flood water would come from if the current cell were to flood. We consider all high resolution cells within the low resolution flooded area as potential sources. We denote the identified source cell as $v'$. Recall that for each high resolution grid cell $v'$ within the low resolution flooded area, we define the predictive distribution for the high resolution flood height as $Y_H(v') \sim Y_D(v') + \hat{\sigma} \mathcal{T}(n-1)$. To get the predictive mean $Y_A(v)$ for $Y_H(v)$ we use the differences between the high resolution elevations at $v$ and $v'$ to shift $Y_D(v')$ as follows:

$$Y_A(v)= [Y_D(v')- (E_H(v)-E_H(v'))]_+.$$
%\begin{equation*}
%    Y_A(v) =
%\left\{
%	\begin{array}{ll}
%		 Y_D(v')- (E_H(v)-E_H(v'))  & \mbox{if } Y_D(v')- (E_H(v)-E_H(v')) \geq 0 \\
%	   0 & \mbox{if } Y_D(v')- (E_H(v)-E_H(v')) < 0
%	\end{array}.
%\right.
%\end{equation*}
Here $E_H(v)$ and $E_H(v')$ are the high resolution elevations at locations $v$ and $v'$. We describe the estimation process for $\hat{\sigma}$ in Subsection 2.4. 

%MH: COULD YOU ADD A HIGH-LEVEL SUMMARY OF THE ENTIRE APPROACH HERE? SOMETHING THAT WOULD FIT INTO LESS THAN HALF A PAGE? 

\subsection{Summary}

We summarize our approach in terms of the information we start with, the quantities we compute, and the resulting distributions. Our process begins with three types of information: a calibrated low (10 m) resolution projection, observed data, and elevation maps at the high (5 m) and low resolution. We start by specifying the model for high resolution flood heights within low resolution wet cells as a shifted and scaled t distribution. First we compute the downscaled flood heights using the low resolution flood heights and the high and low resolution elevation maps. We shift the t distribution using the downscaled flood heights. We set the scaling parameter for the t distribution using the downscaled flood heights and the observed flood heights. We set the number of degrees of freedom in the t distribution equal to the number of observations minus one.

Next we specify our model for high resolution flood heights within low resolution dry cells. We first use the downscaled flood heights within low resolution wet cells and and high resolution elevations to compute the adjusted downscaled flood heights within dry low resolution cells. We then estimate the flooding probability of high resolution cells using the low resolution flood heights and the low and high resolution elevations. 
We model high resolution flood heights within dry low resolution cells using a mixture of a scaled and shifted t distribution and a point mass at zero. The probability of the flood height coming from the t distribution is equal to the estimated probability of the high resolution cell being flooded. We shift this distribution by the adjusted downscaled flood heights and scale it by the same factor as the wet cell model. The probability of the flood height coming from the point mass at zero is equal to the estimated probability of the high resolution cell not being flooded.

\section{Results}\label{Sec: Results}

\subsection{Approximating the high resolution flood projection for Hurricane Ida}\label{SS: Comparison}

PDFlood provides accurate approximations of high resolution flood heights and appropriate uncertainty quantification for the Hurricane Ida test case (Table \ref{Table:Downscaling Performance PDFlood VS CostGrow}). To judge PDFlood's performance, we evaluate (1) how close the mean of our predictive distribution is to the 5 m resolution projection, (2) whether the 5 m resolution flood heights fall within our 95\% prediction intervals, and (3) how well the flooding probabilities given by our approach identify flooded and dry cells. Our approach provides comparable deterministic flood height estimates to the state-of-the-art approach for downscaling riverine flood hazards while quantifying neglected uncertainties. We adopt CostGrow as the state-of-the-art in downscaling riverine flood hazards due to its superior performance to other flood model-specific downscaling approaches \citep{bryant_etal_2024}.

Compared to CostGrow in this case study, PDFlood provides similarly accurate approximate high resolution flood heights and more accurate classifications of high resolution cells as flooded or not flooded (Table \ref{Table:Downscaling Performance PDFlood VS CostGrow}). To compute the mean absolute error (MAE) for PDFlood, we compare the mean of the predictive distribution for each grid cell to the true high resolution flood height. Deterministic approaches like CostGrow predict a single flood height for each high resolution grid cell. 
The MAE of PDFlood is 0.13 m, while the MAE of CostGrow is 0.14 m. We also compare PDFlood to CostGrow in terms of the ability to identify high resolution cells as flooded or not flooded. Following previous literature \citep{schubert_etal_2024}, we call a cell flooded if the flood height is greater 0.3 m. Using PDFlood we classify a cell as flooded if its flooding probability is greater than 0.5, i.e. if $P(Y_H(v)>0.3\textnormal{ m}) >0.5$. CostGrow classifies a cell as flooded if the predicted flood height is greater than 0.3 m. PDFlood accurately classifies 96\% of all high resolution grid cells, while CostGrow accurately classifies 93\% of all high resolution grid cells. 

\begin{table}[h]
\caption{We compare the performance of PDFlood to CostGrow \citep{bryant_etal_2024} in terms of how well downscaled projections approximate 5 m resolution projections. Satisfactory values of mean absolute error, 95\% prediction interval coverage, and percent of flooded and non-flooded cells identified are less than 0.5 m, greater than 90\%, and greater than 90\%, respectively. Bold font indicates the better value of the metric between the two approaches.}
\centering
\begin{tabular}{r|ll}
  \hline
  & PDFlood & CostGrow \\ 
  \hline
  Mean absolute error (m) & \textbf{0.13} & 0.14 \\
  95\% prediction interval coverage & \textbf{98\%} & NA \\
  Percent of flooded and non-flooded cells identified & \textbf{96\%} & 93\% \\
  \hline
\end{tabular}
\label{Table:Downscaling Performance PDFlood VS CostGrow}
\end{table}

\begin{figure}
    \centering
    \includegraphics[angle=270,width=.8\linewidth]{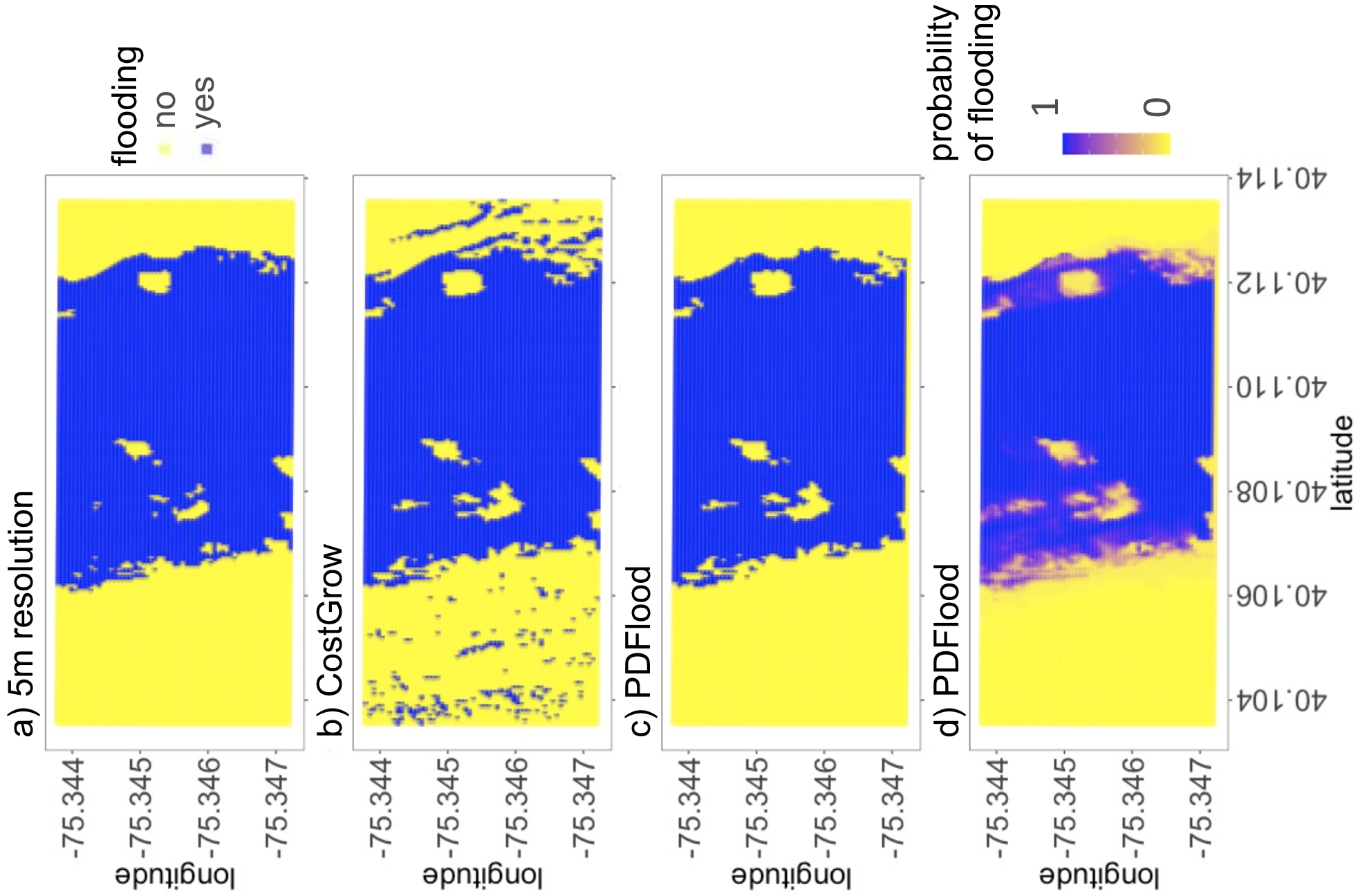}
    \caption{Comparison of flood hazards derived from different methods. We compare our probabilistic downscaling approach to (b) CostGrow \citep{bryant_etal_2024} in terms of ability to predict (a) exposure to at least 30 cm of flooding according to LISFLOOD run at the 5 m resolution. PDFlood provides a (d) probability of exposure to at least 30 cm of flooding for each 5 m resolution grid cell. To obtain (c) we choose cutoff of 0.5 and label all cells with greater than 0.5 probability of exposure as flooded with at least 30 cm.}
    \label{Fig:comparePDFloodToCostGrow}
\end{figure}

PDFlood offers spatial flooding probabilities which are not offered by CostGrow (Figure \ref{Fig:comparePDFloodToCostGrow}). By design, a single implementation of deterministic downscaling is silent on uncertainties. These spatial flooding probabilities quantify uncertainty surrounding whether a cell will flood. For grid cells just outside the region predicted to be flooded, a deterministic downscaling approach only provides a predicted flood height of zero. On the other hand, PDFlood provides flooding probabilities that tend to be close to one half near the boundary of the high resolution flooded region. The uncertainty provided by PDFlood enables decision-makers to consider their degree of safety as opposed to only whether or not they are safe from flooding. %These probabilities are close to one over most of the 5 m resolution flooded region and close to zero over most of the 5 m resolution non-flooded region. 

\begin{figure}
    \centering   
    \includegraphics[width=0.21\linewidth]{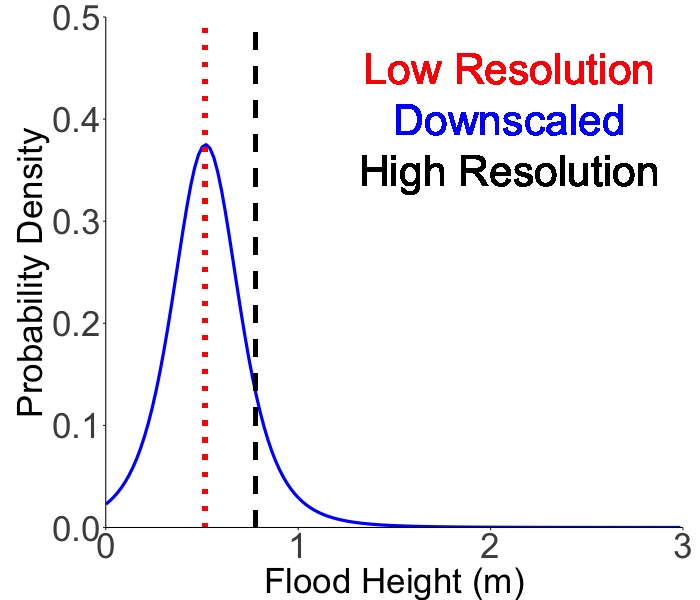}
    \centering
    \includegraphics[width=0.18\linewidth]{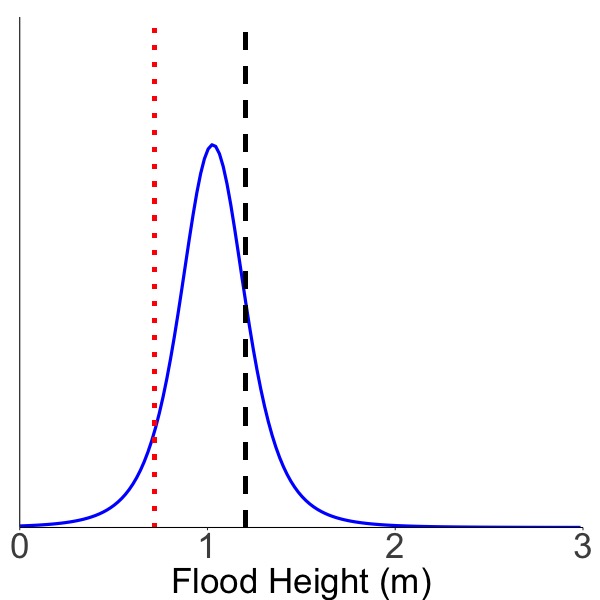}
    \centering
    \includegraphics[width=0.18\linewidth]{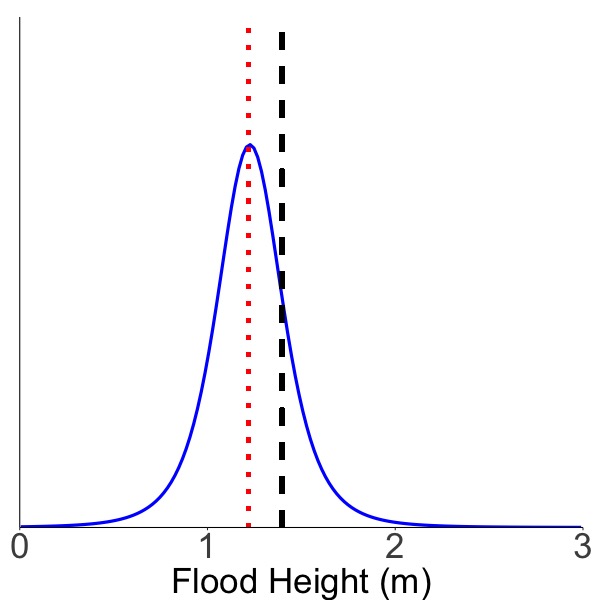}
    \includegraphics[width=0.18\linewidth]{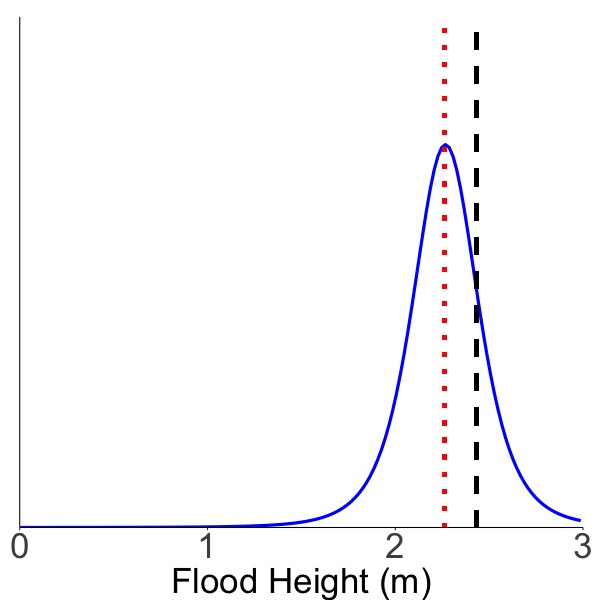}
    \includegraphics[width=0.18\linewidth]{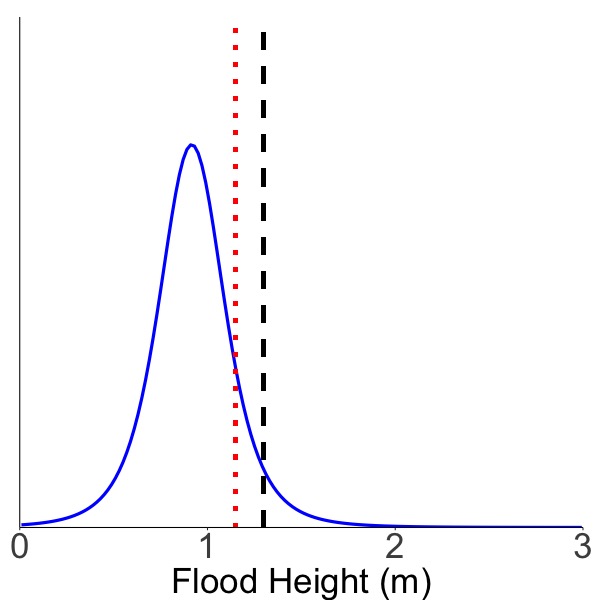}
    \caption{The probability densities of flood heights for the high resolution grid cells containing each of the five observed high water marks from Hurricane Ida \citep{USGS_HWMs}. We compare the high resolution flood height we aim to approximate, the low resolution flood height used for downscaling, and the probability distribution of flood heights provided by our downscaling approach, PDFlood.}
    \label{Fig: example flood height PDFs}
\end{figure}

PDFlood also provides a probability distribution of flood heights for each 5 m resolution grid cell (Figure \ref{Fig: example flood height PDFs}). For illustration, we consider the 5 m resolution grid cells containing each of the five observed high water marks from Hurricane Ida. The bulk of PDFlood's predictive distribution often contains the high resolution flood height. The 98\% empirical coverage of PDFlood's 95\% prediction intervals supports the results of this visual check. The 10 m resolution flood height or the mean of PDFlood's predictive distribution alone provides no estimate of the potential error magnitude, reinforcing the value of a probabilistic downscaling approach.

\subsection{Other flood events}

PDFlood generalizes well to flood events for which no observations are available (Table \ref{Table:Downscaling Performance Other Flood Events}). We consider the flood events Tropical Storm Isaias in 2020 (river discharge: 2503 $\frac{m^3}{s}$), a record-setting rainfall event in 2014 (river discharge: 2560 $\frac{m^3}{s}$), and one hypothetical more extreme future flood events (river discharge: 3681 $\frac{m^3}{s}$) \citep{USGS_peakflow}. In terms of MAE our approach performs slightly better when the river discharge is lower than that of Hurricane Ida and slightly worse when the river discharge is higher than that of Hurricane Ida. This suggests a potential pattern where downscaling is easier for less extreme flood events. However, the 95\% prediction interval coverage of the 5 m resolution flood height stays at 98\% as river discharge increases. The percent of high resolution grid cells accurately classified as flooded or not flooded also stays at 98\% as river discharge increases.

PDFlood performs comparably to CostGrow for these other flood events (Table \ref{Table:Downscaling Performance Other Flood Events}). For the 2020 flood event, the mean absolute errors (MAE) from PDFlood and CostGrow are almost identical with the MAE from PDFlood being slightly lower. For the 2014 flood event, the MAEs from the two approaches are identical. In terms of accurately classifying cells as flooded or not, PDFlood (98\%) slightly outperforms CostGrow (96\%) for these two events. For the hypothetical future flood event CostGrow gives a slightly lower MAE (0.14 m) compared to PDFlood (0.16 m). The two approaches yield identical flooded-versus-non-flooded classification accuracies (98\%) for this event. For all flood events PDFlood provides prediction intervals with near-theoretical coverage which are neglected by CostGrow.

\begin{table}[h]
\caption{Evaluating the performance of PDFlood compared to CostGrow in terms of how well downscaled projections approximate 5 m resolution projections for two historical and one potential future flood event. Satisfactory values of mean absolute error, 95\% prediction interval coverage, and percent of flooded and non-flooded cells identified are less than 0.5 m, greater than 90\%, and greater than 90\%, respectively. Bold font indicates a better value of the metric.}
\centering
\begin{tabular}{rllll}
  \hline
   &  PDFlood & CostGrow \\ 
  \hline
  2014 flood event \\
  \hline
  Mean absolute error (m) & 0.10 & 0.10 \\
  95\% prediction interval coverage & \textbf{98\%} & NA \\ 
  Flooded and non-flooded cells correctly identified &  \textbf{98\%} & 96\% \\
  \hline
  2020 flood event\\
  \hline
  Mean absolute error (m) & \textbf{0.096} & 0.10 \\
  95\% prediction interval coverage & \textbf{98\%}& NA \\ 
  Flooded and non-flooded cells correctly identified &  \textbf{98\%} & 96\% \\
  \hline
  Possible future flood event \\
  \hline
  Mean absolute error (m) & 0.16 & \textbf{0.14} \\
  95\% prediction interval coverage & \textbf{98\%} & NA \\ 
  Flooded and non-flooded cells correctly identified &  98\% & 98\% \\
  \hline
\end{tabular}
\label{Table:Downscaling Performance Other Flood Events}
\end{table}

\subsection{Computational Time Savings}

Our approach is comparably very fast relative to the time needed to obtain a high resolution flood projection. With a starting resolution of 10 m our downscaling approach takes 3.7 minutes of wall time. On average, obtaining a 5 m resolution projection takes 33 times longer and can take as much as 115 times longer than running PDFlood (Table \ref{Table:TimeToGetRuns}). CostGrow only takes 5 seconds of compute time. However, for some applications the uncertainty quantification provided by PDFlood can well be worth the additional 3.6 minutes. Identifying the cell from which flood water would flow to each high resolution cell outside the low resolution flooded area is by far the most computationally costly step of PDFlood, and this step can be parallelized to further reduce computational costs.

\section{Discussion}\label{Sec: Probabilistic Downscaling Discussion}

We develop PDFlood, a probabilistic flood hazard downscaling approach. PDFlood combines flood model-specific downscaling techniques with model-based uncertainty quantification. Similar to the state-of-the-art in flood model-specific downscaling \citep{bryant_etal_2024}, PDFlood utilizes cost distance analysis to downscale flooded and non-flooded low resolution cells differently, exploiting high resolution elevation data. Similarly to other statistical downscaling approaches, PDFlood uses observed flood heights to specify a probability distribution for high resolution flood heights.

PDFlood improves in this case study on the current state-of-the-art in downscaling flood hazards \citep{bryant_etal_2024} through comparable accuracy and improved uncertainty quantification. For all considered flood events, PDFlood gives comparably accurate point estimates of high resolution flood heights. PDFlood also yields a similar error rate in classifying high resolution grid cells as flooded or not. More importantly, PDFlood supplies previously neglected uncertainty estimates for both flood height estimates and classifications of flooded versus non-flooded. The uncertainties for flood height estimates come in the form of prediction intervals, and the uncertainties for classifications of flooded versus non-flooded come in the form of flooding probabilities. Decision makers can use these prediction intervals to better protect their property against uncertain extreme flood events.

Considering uncertainties improve decisions on how to manage flood risks \citep[e.g.][]{zarekarizi_2020}. Here we introduce an approach to account for uncertainties due to downscaling. %We hope that PDFlood users who intend to make flood risk management decisions will consider other relevant forms of uncertainty. These include uncertainties surrounding flood hazards, building characteristics, and the discount rate \citep{zarekarizi_2020}. Neglecting parametric uncertainty can lead to underestimation of extreme flood events \citep{sanjib_etal_2023}. Considering deep uncertainties surrounding flood hazards, building characteristics, and the discount rate can lead to more protective management decisions \citep{zarekarizi_2020}. 

%Consider a homeowner who is debating whether or not to elevate their home. A deterministic downscaling approach might project that they will experience no flooding for a flood event with a 20-year return period. However, PDFlood might project a 40\% chance of flooding for this event. If the homeowner only has access to the flood projection from the deterministic downscaling approach, they likely would not choose to take any measures to reduce their home's vulnerability. However if they have access to a downscaled projection from PDFlood, they may consider taking steps to reduce their home's vulnerability. No matter the choice they make the uncertainty estimates provided by PDFlood allow them to make a more informed decision.

\subsection{Caveats}

We design PDFlood to be relatively simple and fast. We simplify our model structure compared to other statistical downscaling approaches and model flood heights according to a t distribution to cope with the few available high water marks typical of flood hazard modeling applications. However, our estimator for $\sigma^2$ is likely an overestimate due to our use of observations in place of high resolution flood heights in our calculation. 

Our study design also faces certain limitations. First, we configure LISFLOOD-FP by treating channel roughness and floodplain roughness as spatially constant. For the small reach of the Schuylkill river we focus on, we find that our assumption of spatially constant channel roughness is reasonable. In addition, previous studies show that flood hazard projections from LISFLOOD-FP are rather insensitive to floodplain roughness \citep{me2023,iman_etal_2024}. Second, we downscale flood projections that are already a relatively high resolution. Even so, the computational speedup of using downscaled 10 m resolution projections compare to 5 m resolution projections is large. We choose 10 m resolution projections to due to our limited ability to calibrate lower resolution projections using few high water marks. Third, we test PDFlood using a relatively small area so that we could evaluate PDFlood's performance against the high resolution projections. PDFlood may be more useful for larger areas where we cannot obtain high resolution projections for performance evaluation. In future research we plan to focus on downscaling lower spatial resolutions over larger areas and vary parameters spatially. In our case study we focus on just one flood type: riverine flooding. PDFlood may be useful for other types of flooding which come from one direction, such as storm surge or a levee break.

\begin{acks}[Acknowledgments]
We are grateful to Sara Santamaria Aguilar, Adam Pollack, Arthur Moreno, Milo Moreno, Lilikoy Rothwell, and Skip Wishbone for their valuable inputs. 
\end{acks}
%%%%%%%%%%%%%%%%%%%%%%%%%%%%%%%%%%%%%%%%%%%%%%
%% Funding information, if any,             %%
%% should be provided in the                %%
%% funding section.                         %%
%%%%%%%%%%%%%%%%%%%%%%%%%%%%%%%%%%%%%%%%%%%%%%
\begin{funding}
This work was supported by the U.S. Department of Energy, Office of Science, Biological and Environmental Research Program, Earth and Environmental Systems Modeling, MultiSector Dynamics under cooperative agreement DE-SC0022141. Additional support was provided by the Penn State Center for Climate Risk Management and the Thayer School of Engineering at Dartmouth College. Any opinions, findings, and conclusions or recommendations expressed in this material are those of the authors and do not necessarily reflect the views of the US Department of Energy or other funding entities. Computations for this research were performed on the Pennsylvania State University’s Institute for Computational and Data Sciences’ Roar supercomputer.
\end{funding}

%%%%%%%%%%%%%%%%%%%%%%%%%%%%%%%%%%%%%%%%%%%%%%
%% Supplementary Material, including data   %%
%% sets and code, should be provided in     %%
%% {supplement} environment with title      %%
%% and short description. It cannot be      %%
%% available exclusively as external link.  %%
%% All Supplementary Material must be       %%
%% available to the reader on Project       %%
%% Euclid with the published article.       %%
%%%%%%%%%%%%%%%%%%%%%%%%%%%%%%%%%%%%%%%%%%%%%%

\section{Supplement}
%\stitle{Supplementary Information Supporting Probabilistic Downscaling for Flood Hazard Models}
%\sdescription{???.}

\subsection{Details on Bayesian emulation-calibration}\label{SubSec:EmulationCalibration}
In calibration we infer computer model parameters by comparing the associated outputs to observational data \citep[e.g.][]{kennedy2001bayesian}. In our case the computer model outputs are spatial flood projections. A flood projection is a collection of flood heights at all locations for a given parameter setting. The observational data are simulated flood heights at the 3 m resolution at the locations of local businesses. The computer model calibration framework accounts for sources of uncertainty stemming from uncertain input parameters, observational errors, and model-observation discrepancies. Because we simulate observations at a higher spatial resolution than the flood projections used for calibration, we still need to estimate the model-observation discrepancy.

We define $Y(\theta,s)$ as the computer model output at location $s \in \mcS$ and parameter setting $\theta \in \Theta$. In our case $\mcS \subset \mbR^2$ is the spatial domain of interest consisting of latitude and longitude pairs. We denote the $m$ locations defining the center of each grid cell of the computer model output as $\bs^M= (s^M_1,...,s^M_m)$. Since we only infer channel roughness, the parameter space $\Theta = (0.01,0.1) \subset \mbR^1$. %$\bZ(s)$ is the observation at location $s$. 
We have access to computer model runs at $p$ design points $\btheta= (\theta_1,...,\theta_p)$. Each flood projection $\bY(\theta_i,\bs^M)=$ $(Y(\theta_i,s^M_1),...,  Y(\theta_i,s^M_m))^T$ $\in \mbR^{m}$ at parameter setting $\theta_i$ is a spatial process. The observations $\bZ= (Z(s^O_1),..., Z(s^O_n))^T \in \mbR^{n}$ occur at locations of local businesses $\bs^O= (s^O_1,...,s^O_n)$. %for $\{s_{k_1},..., s_{k_n}\} \subset \{s_1,...,s_m\} $ is the vector of observations. 
We treat the computer model as a `black box.' In other words, we only change the model inputs. 

The Bayesian calibration framework calls for a computer model run at each iteration of the MCMC algorithm. This can require tens of thousands of sequential iterations. This approach can be computationally prohibitive for computer models with even moderately long run times on the order of minutes. This includes the 10 m version of LISFLOOD-FP which would take about 41 days of wall time to produce 10000 model evaluations. We mitigate this problem by constructing a Gaussian process emulator based on a training set of computer model runs. 

% Need to add basis for Computer Model Calibration %
%\subsection{3.2 Two-stage Emulation and Calibration Method}\label{2stage EC}
We calibrate LISFLOOD using a two-stage emulation-calibration approach. We first fit an emulator to the computer model (Equation \eqref{EQ:GPemulationCh3}) and then calibrate the resulting emulator using observations (Equation \eqref{EQ:GPcalibrationCh3}). Single-stage methods \citep{higdon2008} combine the two stages (emulation and calibration) into a single inferential step. However, two-stage approaches have several advantages over single-stage methods. Fitting the emulator in a separate step using only computer model output prevents contamination by the calibration model parameters \citep[e.g.][]{liu_etal_2009, bhat_haran_goes2010,bayarri2007framework}. This allows for easier emulator evaluation and reduces identifiability issues. Two stage emulation calibration is also more computationally efficient.

We provide an overview of the two-stage emulation-calibration framework for a spatial computer model. We also provide details of how we apply this framework in our study. In the emulation step we fit a Gaussian process emulator to training data. The training data is comprised of concatenated spatial computer model outputs $\mathbf{Y}= \big(\bY(\theta_{1},\bs^O),...,\bY(\theta_{p},\bs^O)\big)^T$ evaluated at the $p$ parameter settings and $n$ observation locations. 
We construct the Gaussian process emulator $\bfeta(\theta,\bY)$ by fitting the model:
\begin{align}\label{EQ:GPemulationCh3}
    \bY \sim \mcN(h(\bX)\bbe,\Sigma_{\bxi}(\bX) + \sigma^2 \bI).
\end{align}
\noindent Here $\bX$ is a $np \times b$ matrix of covariates including the spatial locations and computer model parameter settings. $h(\bX)$ is a function $h(\cdot)$ of the matrix $\bX$. We also use $\bX$ to compute the covariance matrix $\Sigma_{\bxi}(\bX)$. We estimate the emulator parameters including  the vector of covariance parameters $\bxi$, the regression coefficients $\bbe \in \mbR^b$, and the nugget parameter $\sigma^2 \in \mbR$. We estimate $\bbe$, $\bxi$, and $\sigma^2$ by maximizing the likelihood to fit a Gaussian random field to $\bY$. This field gives a probability model for the computer model run at any parameter setting $\theta \in \Theta$ and any location $s \in \mcS$. The Gaussian process model gives a predictive distribution for $\bY(\theta^0,s^0)$ at unobserved $\theta^0$ and $s^0$ given the runs used for training $\bY$. The emulator $\bfeta(\bY,\theta)$ is the resulting interpolated process. 

In our study we centered $\bY$ before emulation and set $h(\bX)=0$, eliminating the need to estimate $\bbe$. We fit a Gaussian process emulator at each observation location independently, because we only need computer model output approximated at the observation locations. We fit the Gaussian processes in parallel using an unconstrained quasi-Newton Method optimizer \citep{PORT} to maximize the likelihood. Our approach results in the following $n$ emulators: $$ \bY(\btheta,s^O_i) \sim \mcN(\bzero, \Sigma_{\bxi_i}(\btheta) + \sigma^2_i \bI)\textnormal{, for } i=1,...,n.$$
In our case, $\bfeta(\bY,\theta)$ is the collection of independent emulators $\bfeta(\bY(s^O_i),\theta)$ at all observation locations $\bs^O$. $\bfeta(\bY,\theta)$ gives a predictive distribution for $\bY(\theta^0,s^O_i)$ at unobserved $\theta^0$ and any $s^O_i \in \bs^O$ given the runs used for training at the same location $\bY(\btheta,s^O_i)$.

In the calibration step, we model the observed data $\bZ$ with respect to the emulator $\eta(\bY,\theta)$ as follows:
\begin{align} \label{EQ:GPcalibrationCh3}
    \bZ &= \eta(\bY,\theta) + \bdel + \bep \nonumber, \\
    \bdel &\sim \mcN(\bzero,\bSig_{\bga}(\bs)), \textnormal{ and} \\
    \bep &\sim \mcN(\bzero,\sigma^2_\epsilon \bI). \nonumber
\end{align}
The discrepancy term $\bdel$ captures the systematic differences between emulator projections and observations \citep[e.g.][]{chang2014}. $\bSig_{\bga}(\bs)$ is a spatial covariance matrix. $\bep$ represents the independent and identically distributed observational errors \citep[e.g.][]{chang2014}. We typically infer $\theta$, $\bdel$ and $\sigma^2_\epsilon$ by sampling from the posterior distribution $\pi(\theta,\bdel,\sigma^2_\epsilon|\bZ)$ via MCMC. 

In our study we use the following simplified approach to estimate $\bdel$. First, we find the mean absolute error (MAE) between $\bY(\theta_j,\bs^O)$ and $\bZ$ for all $j= 1,...,p$. For each observation location $s^O_i$, we compute the average difference between $\bY(\theta_j,s^O_i)$ and $\bZ(s^O_i)$ over the three $\theta_j$s with the smallest MAEs. We denote the vector of average differences $\hat{\bdel} = (\hat{\delta}(s^O_1),...,\hat{\delta}(s^O_n))^T$ and use it in place of $\bdel$ in calibration. We use this simplification to simplify calibration and save computational costs. We then take the posterior mean of $\theta$, $\theta^*$, to be a point estimate of the channel roughness value that gives us the low resolution flood projection most similar to the observational data. We find that our simplified calibration procedure is able to closely replicate the observational data considering the difference in resolution between our observation and our model projections. Thus our approach neglects uncertainty surrounding $\btheta^*$ and $\bdel$.

\subsection{Details on modeling flooding probability with elevation}\label{S: details flood prob elev}

\begin{figure}
    \centering
    \includegraphics[width=.55\linewidth]{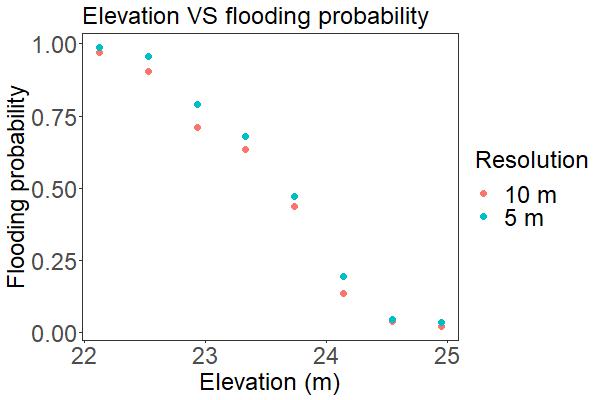}
    \caption{We plot midpoint of each elevation group (x axis) against the proportion of cells that are flooded in said elevation group (y axis). The relationship between elevation and flooding probability is similar whether the cells are 10 m or 5 m resolution.}
    \label{Fig: elev vs flood prob}
\end{figure}

We produce Figure \ref{Fig: elev vs flood prob} by repeating these steps for each spatial resolution. First we identify the minimum elevation for which any 10 m resolution cell is not flooded (21.9 m) and the maximum elevation for which any 10 m resolution cell is flooded (25.1 m). We then divide the cells with elevations between these two values into eight groups so that each group has an elevation range of 0.4 m. For each group, we compute the proportion of cells that are flooded. Figure \ref{Fig: elev vs flood prob} plots the midpoint value of each elevation group against the proportion of cells that are flooded in that elevation group.

\subsection{Additional results}

\begin{figure}
    \centering
    \includegraphics[width=0.5\linewidth]{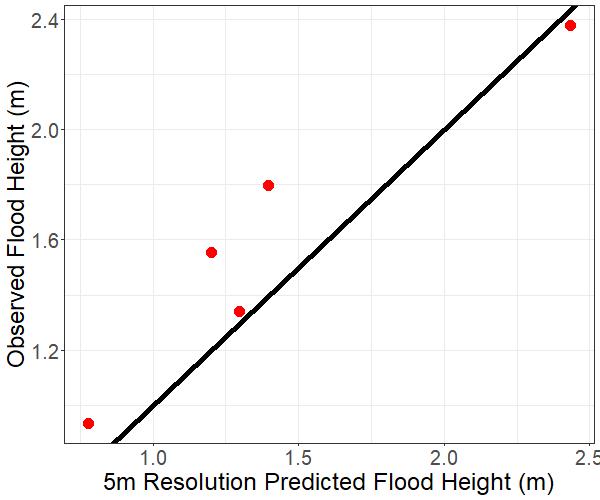}
    \caption{We plot the predicted flood height according to the calibrated 5m resolution model at the observation locations (x axis) against the observations, i.e. high water marks (y axis). For reference, we display a black line representing what perfect model performance would look like.}
    \label{Fig: 5m Preds VS HWMs Supplement}
\end{figure}

The calibrated projections from LISFLOOD-FP run at the high spatial resolution (5 m) are similar to the recorded high water marks for Hurricane Ida (Figure S\ref{Fig: 5m Preds VS HWMs Supplement}). The calibrated high resolution projection slightly underestimates four out of five of the high water marks and slightly overestimates the deepest high water mark. Given the scale of the data and the limited availability of high water marks, we deem the 5 m resolution model performance (MAE = 0.2 m) to be satisfactory.

CostGrow performs slightly better in terms of accurately identifying flooded cells, but PDFlood better in terms of accurately identifying non-flooded cells (Table S\ref{Table: PDFlood VS CostGrow Supplement}). The difference in performance on non-flooded cells is greater than the difference in performance on flooded cells. PDFlood accurately classifies 93\% of flooded high resolution grid cells. CostGrow classifies 95\% of flooded high resolution grid cells. However, PDFlood accurately classifies 99.9\% of non-flooded high resolution grid cells, and CostGrow accurately classifies 92\% of non-flooded high resolution grid cells. 

\begin{table}[b]
\caption{Comparing the performance of PDFlood to CostGrow in terms of how well downscaled projections approximate 5 m resolution projections. Satisfactory values of percent of flooded cells identified and percent of non-flooded cells identified are both greater than 90\%. Bold font indicates the better value of the metric between the two approaches. * indicates that the value has been rounded up to 100\%.}
\centering
\begin{tabular}{r|ll}
  \hline
  & PDFlood & CostGrow \\ 
  \hline
  Percent of flooded cells identified &  93\% & \textbf{95\%} \\ 
  Percent of non-flooded cells identified & \textbf{100\%}* & 92\% \\ %.99
  \hline
\end{tabular}
\label{Table: PDFlood VS CostGrow Supplement}
\end{table}

For the other flood events considered, PDFlood still correctly identifies a very high percentage of both flooded and non-flooded cells (Table S\ref{Table: PDFlood Other Flood Events Supplement}). PDFlood correctly identifies practically all non-flooded cells up to a rounding error. PDFlood also correctly identifies at least 95\% of flooded cells for all other flood events considered. PDFlood's performance on flood events that we do not have high water marks for is at least as good as its performance on Hurricane Ida in terms of these metrics. Compared to CostGrow, PDFlood is still slightly better at identifying non-flooded cells and slightly worse at identifying flooded cells. However, CostGrow never identifies more than 2\% more of the flooded cells, and PDFlood consistently identifies 4\% more of the non-flooded cells.

\begin{table}[t]
\caption{Evaluating the performance of PDFlood compared to CostGrow in terms of how well downscaled projections approximate 5 m resolution projections for two historical and one potential future flood event. Satisfactory values of percent of flooded cells identified and percent of non-flooded cells identified are both greater than 90\%. Bold font indicates the better value of the metric between the two approaches. * indicates that the value has been rounded up to 100\%.}
\centering
\begin{tabular}{rllll}
  \hline
   &  PDFlood & CostGrow \\ 
  \hline
  2014 flood event \\
  \hline
  Flooded cells correctly identified & 95\% & \textbf{97\%} \\ 
  Non-flooded cells correctly identified &  \textbf{100\%}* & 96\% \\
  \hline
  2020 flood event\\
  \hline
  Flooded cells correctly identified & 95\% & \textbf{97\%} \\ 
  Non-flooded cells correctly identified &  \textbf{100\%}* & 96\% \\
  \hline
  Possible future flood event \\
  \hline
  Flooded cells correctly identified & 97\% & \textbf{99\%} \\ 
  Non-flooded cells correctly identified &  \textbf{100\%}* & 96\% \\
  \hline
\end{tabular}
\label{Table: PDFlood Other Flood Events Supplement}
\end{table}

%% if your bibliography is in bibtex format, uncomment commands:
\bibliographystyle{imsart-nameyear} % Style BST file
\bibliography{main}       % Bibliography file (usually '*.bib')

%% or include bibliography directly:
% \begin{thebibliography}{}
% \bibitem[\protect\citeauthoryear{???}{???}]{b1}
% \end{thebibliography}

\end{document}